\documentstyle[bbm,amsfonts,epsfig,12pt,here]{article}

\newcommand {\eqref} [1] {(\ref {#1})}
\newcommand {\slsh} [1] {\not{\hbox{\kern-2pt${#1}$}}}

\def\drawbox#1#2{\hrule height#2pt
        \hbox{\vrule width#2pt height#1pt \kern#1pt
              \vrule width#2pt}
              \hrule height#2pt}

\def\Asym#1#2{\vcenter{\vbox{\drawbox{#1}{#2}
              \kern-#2pt       
              \drawbox{#1}{#2}}}}

\def\asymm{\Asym{6.4}{0.3}}

\def\basymm{\overline{\asymm}}

\newcommand {\beq} {\begin{equation}}
\newcommand {\eeq} {\end{equation}}
 \newcommand {\ber}{\begin{eqnarray*}}
 \newcommand {\eer} {\end{eqnarray*}}
\newcommand {\bea}{\begin{eqnarray}}
 \newcommand {\eea} {\end{eqnarray}}
\newcommand{\Nfour} {${\cal N}=4\ $}

\newcommand{\None}{${\cal N}=1\ $}
\newcommand{\ztwo}{${\bf Z}_2\ $}

\def\Acknowledgements{\bigskip  \bigskip {\begin{center} \begin{large}
             \bf ACKNOWLEDGMENTS \end{large}\end{center}}}

\begin{document}
\begin{titlepage}
\begin{flushright}{CERN-TH/2003-057

TPI-MINN-03/08, UMN-TH-2133/03
}
\end{flushright}
\vskip 0.8cm

\centerline{{\Large \bf The Cosmological Constant and Domain Walls in}}
\vskip 0.1cm
\centerline{{\Large \bf Orientifold Field Theories and  \boldmath{\None} Gluodynamics}}
\vskip 1cm
\centerline{\large A. Armoni ${}^a$ and M. Shifman ${}^{a,b}$}
\vskip 0.1cm
\centerline{\small adi.armoni, michael.shifman@cern.ch}
\vskip 0.4cm
\centerline{${}^a$ Theory Division, CERN}
\centerline{CH-1211 Geneva 23, Switzerland}
\vskip 0.4cm
\centerline{${}^b$ William I. Fine Theoretical Physics Institute, University
of Minnesota,}
\centerline{Minneapolis, MN 55455, USA$^\star$}
\vskip 0.8cm

\begin{abstract}

We discuss domain walls and vacuum energy density
(cosmological constant)  in \None gluodynamics and in
 non-supersymmetric large $N$ orientifold
field theories which have been
recently shown to be planar equivalent
(in the boson sector)  to \None gluodynamics.
A relation between the vanishing force
between two parallel walls and vanishing cosmological constant
is pointed out. This relation may
 explain why 
 the cosmological constant vanishes  in the orientifold field theory
at leading order although the hadronic spectrum of this theory
does not contain fermions  in the limit $N\to\infty$. 
The cancellation is among
even and odd parity bosonic contributions, due to NS-NS and R-R
cancellations in the annulus amplitude of the underlying string theory.
We  use the open-closed string channel duality to describe interaction
between the domain walls which is interpreted
as the exchange of composite ``dilatons'' and ``axions''
coupled to the walls.
Finally, we study some planar equivalent pairs
in which both theories in the parent-daughter pair are
non-supersymmetric.

\end{abstract}

\vspace{0.5cm}

\noindent
\rule{2.4in}{0.25mm}\\
$^\star$ Permanent address.
\end{titlepage}

$$
$$

\section{Introduction}

\noindent

Domain walls are BPS objects which appear in \None supersymmetric
(SUSY) gluodynamics \cite{Dvali:1996xe}. If the gauge group is
SU($N$), there are $N$ distinct discrete  vacua labeled by the order parameter,
the gluino condensate,
\beq
 \langle \lambda \lambda \rangle_k = N \Lambda ^3 \exp\left( i {2\pi k\over N}
\right)\,,\qquad k=0,1,2,..., N-1.
 \label{condensate}
\eeq 
The domain wall $W_{\{k,k+1\}}$ interpolates
between the $k$-th and $k+1$ vacua.
Moreover, at $N\to\infty$ two parallel domain walls
$W_{\{k,k+1\}}$ and $W_{\{k+1,k+2\}}$
are also BPS --- there is neither attraction nor repulsion between them.

\vspace{1mm}

It is known that the BPS domain walls  in \None  gluodynamics
present a close parallel  to D branes in string
 theory \cite{Witten:1997ep}. In particular,
a fundamental flux tube can end on the BPS domain wall, similarly
to F1 ending on D branes in string theory \cite{Witten:1997ep}
(see also 
\cite{Kogan:1997dt,Gabadadze:1999pp,Polchinski:2000uf,Klebanov:2000hb,Loewy:2001pq}). 
The purpose of this paper is three-fold.
First, we show that this parallel can be further extended.
In string theory the absence of forces between parallel D branes is due to 
a cancellation between the interactions induced by NS-NS and R-R charges.
We show how a similar cancellation works in  \None  gluodynamics.
This parallel yields an important insight
revealing a relation between the vanishing of the cosmological
constant and  cancellation of forces. This observation will be used
later.

\vspace{1mm}

Second, we will extend this parallel to
a {\em  non-supersymmetric} gauge field theory.
Recently, we discussed a non-supersymmetric theory, where exact results
on the strong coupling regime could be obtained \cite{Armoni:2003gp}.
 The theory, named
``orientifold field theory,'' is a daughter of \None gluodynamics.
The parent-daughter relationship is understood 
 in the sense
of \cite{Strassler:2001fs}. The parent theory is \None gluodynamics
with the gauge group U($N$). The daughter theory also has
 U$(N)$ gauge group, the same gauge coupling as the parent one,
and the fermion sector consisting of one
 Dirac fermion in the antisymmetric tensor representation.

\vspace{1mm}

The advantage of the orientifold daughter
compared to  orbifold  discussed by
Strassler  \cite{Strassler:2001fs} is the absence of the
twisted sector in the former. The {\em nonperturbative} planar
equivalence between   \None gluodynamics and its orbifold,
conjectured by Strassler, was questioned in the 
literature (see e.g. Refs.~\cite{Gorsky:2002wt,Tong:2002vp}),
with the twisted sector of the orbifold theory being the main suspect.
The  nonperturbative planar
equivalence between   \None gluodynamics and its orientifold
daughter was shown \cite{Armoni:2003gp} to be on a much more solid
theoretical footing.
We argued that 
the orientifold gauge theory, at large $N$, contains $N$ degenerate vacua, has
a bifermion condensate which serves as 
an  order parameter, much in the same way as
the gluino condensate, Eq. (\ref{condensate}).
Another  finding was the vanishing
of the cosmological constant at order $N^2$. These results
seem to be  surprising since the hadronic spectrum of the 
orientifold theory is purely
bosonic.

\vspace{1mm}

The orientifold theory has $N$ discrete degenerate vacua. Hence,
one can expect
domain walls. Indeed, the daughter inherits 
domain walls from its supersymmetric parent.
Two parallel  walls of the type
$W_{\{k,k+1\}}$ and $W_{\{k+1,k+2\}}$
are at indefinite equilibrium. In this sense they are
``BPS,'' although the standard definition of
``BPS-ness,'' through central charges and supercharges,
is certainly not applicable in the non-supersymmetric theory.
In this paper we elaborate on  physics of the domain walls
 in the orientifold theory.
We will show that these walls carry charges similar to NS-NS and
 R-R charges. In addition, we will argue that an open-closed
string channel duality holds for the analogous field theory
 annulus amplitude. Moreover, by exploiting the similarity
 between string theory and field theory we will provide a reason 
 why the cosmological constant of the gauge theory is zero at order $N^2$
despite the fact
 that the hadronic spectrum of the theory  contains only bosons.

\vspace{1mm}

Finally, in the third part
we explain how the parent-daughter relationship
(nonperturbative planar equivalence)
between \None gluodynamics and its orientifold
can be extended to include pairs of theories none of which is supersymmetric.

\section{ The Orientifold Field Theory}
\label{OFT}

\noindent

This ``orientifold field theory''
 was suggested in Refs. \cite{Armoni:1999gc,Angelantonj:1999qg}
in a somewhat different context. 
The field content of the orientifold gauge field theory  differs from
the one of its
 parent theory, U($N$) SUSY gluodynamics, in that the gluinos are replaced by
one massless Dirac fermions in the rank-two antisymmetric tensor representation of
 U($N$) (denoted by $\asymm + \basymm$). The total number of (say) left-handed 
fermions is thus $N(N-1)$ in the
daughter theory and $N^2$ in the parent theory, which agrees to leading order
in $1/N$. 
The  realization of the
orientifold field theory in 
string theory is as follows:
this theory lives on a brane configuration of type 0A
string theory \cite{Armoni:1999gc} which consists of NS5 branes,
D4 branes and an
orientifold plane --- hence the name ``orientifold field theory.''

\vspace{1mm}

 The massless open strings on the brane
correspond to the ultraviolet (UV) degrees of freedom of the field theory:
the gauge field and the antisymmetric fermion.

\vspace{1mm}

We will assume that our gauge theory has a string theory dual in the
spirit of Ref.~\cite{Maldacena:1997re} (yet to be found, though). It is 
presumably of the type 0B on a curved background, similarly to the
orientifold field theory analog
of \Nfour SYM which is type 0B on $AdS_5 \times RP ^5$
\cite{Angelantonj:1999qg}. In this picture
the closed strings correspond
to the infrared (IR) degrees of freedom,  the glueballs and ``quarkonia.''
 Indeed, the type 0 (closed) strings are purely
bosonic, in agreement with our expectation from the confining
orientifold field theory.
Moreover, the bosonic IR
  spectrum of the gauge theory is even/odd parity degenerate, in 
 accordance with degeneracies between the NS-NS and R-R towers
 of the type 0 string.

\vspace{1mm}

\section{Parallel Domain Walls versus D Branes}
\label{PDWDB}

\noindent 

As was mentioned, the gauge theory fundamental flux tubes
 can end on a BPS domain wall. Let
 us assume that for \None gluodynamics/orientifold theory 
 the domain walls have a realization in terms of D$p$ branes ($p>1$) of the
 corresponding
 type IIB/0B string theory. 
Their world volume
 is $012 + (p-2)$ directions transverse to the 
four-dimensional  space-time 
$0123$.
Specific AdS/CFT
 realizations of domain walls in \None theories are given in Refs.
 \cite{Polchinski:2000uf,Klebanov:2000hb,Loewy:2001pq,Maldacena:2000yy},
 mostly in terms of wrapped D5 branes. We deliberately
do not specify which particular
 branes are used to model the BPS walls, since we  do not
perform actual
 AdS/CFT calculations. 
 D branes carry the NS-NS charge, as well
 as the R-R charge \cite{Polchinski:1995mt}. 
Moreover, interactions induced by these charges exactly cancel
guaranteeing that the parallel D branes neither attract nor repel
each other.

\vspace{1mm}

Let us see how this is realized in \None SYM.  
We start from two parallel BPS walls  $W_{\{k,k+1\}}$ and $W_{\{k+1,k+2\}}$.
Each of them is BPS, with the tension~\cite{Dvali:1996xe}
\beq
T_{\{k,k+1\}} = T_{\{k+1,k+2\}} = \frac{N}{8\pi^2}\, | \langle \mbox{tr} \lambda\lambda
\rangle | \,2\, \sin\frac{\pi}{N}\,.
\label{tension}
\eeq
The tension of the configuration $W_{\{k,k+2\}}$ is
\beq
T_{\{k,k+2\}} = \frac{N}{8\pi^2}\, | \langle \mbox{tr} \lambda\lambda
\rangle | \, 2 \, \sin\frac{2\pi}{N}\,.
\eeq
This means that at leading order in $N$ (i.e.$\,\, N^1$)
two parallel walls  $W_{\{k,k+1\}}$ and $W_{\{k+1,k+2\}}$
do not interact. There is no interaction at the level
$N^0$ either. An attraction emerges at  the level $N^{-1}$.
That the inter-wall interaction potential is $O(N^{-1})$
can be shown on general kinematic grounds
(Ref. \cite{adam} presents a detailed discussion, see Eq. (37);
see also Ref.~\cite{RSV01}. This and other aspects of dynamics of inter-wall
separation will be considered in Ref.~\cite{RSV}.)

\vspace{1mm}

Our task is to understand dynamics
of this phenomenon from the
field theory side. Assume that two walls under
consideration are separated by a distance $Z \gg m^{-1}$
where $m$ is the mass of the lightest composite meson. 
What is the origin of  the force between these walls?

\vspace{1mm}

The interaction is due to the meson exchange in the bulk.
Consider  the lightest mesons, scalar and pseudoscalar.
The scalar meson $\sigma$, the ``dilaton,'' is coupled 
to the trace of the energy-momentum
tensor $\theta^\mu_\mu$,
\beq
\frac{\sigma}{f}\,\theta^\mu_\mu \,  =
  \frac{3N}{16\pi^2 f}\,\, \sigma\, \mbox{tr}\,  F^2 \,,
\label{anomc}
\eeq
(see e.g.  Ref.~\cite{Migdal:jp}).
Here $f$ is a coupling constant scaling as
\beq
f\sim N\, \Lambda\,.
\label{ccs}
\eeq
Integrating over the transverse direction and using the
fact that $\theta^\mu_\mu$ translates into mass, we find that the
``dilaton''-wall coupling (per unit area)
is 
\beq
T\,\sigma/f\,,
\label{odin}
\eeq
 where $T$ is the tension, see Eq. (\ref{tension}).
The $\sigma$-wall coupling  scales as $N^0$.

\vspace{1mm}

The coupling of the pseudoscalar meson
(the ``axion'' or ``$\eta '$''; we will denote this field by $\eta$ ---
it will have a realization in terms of the RR 0-form of type IIB/0B)
to the wall is related to the change of the phase of the gluino condensate
across the wall. 
It can be estimated as~\footnote{Equation (\ref{dva})
guarantees, automatically, that the ``axion''-wall coupling
is saturated inside the wall. Outside the wall, in the vacuum,
 $\alpha =$ const.,
while $\langle  \mbox{tr}\,  F^2\rangle = \langle  \mbox{tr}\,  F\tilde{F}\rangle=0 $. }
\beq
\frac{\eta}{f}\,\int dz\,\,  \frac{N}{8\pi^2}\,\,  \mbox{tr}\,  F\tilde{F}
\to 
\frac{\eta}{f}\,\int dz\,\, \frac{N}{8\pi^2}\, |\,\mbox{tr}\,\lambda\lambda\, |\,\,
\frac{\partial\alpha}{\partial z}\,,
\label{dva}
\eeq
where $z$ is the coordinate transversal to the wall,
and $\alpha$ is the phase of the order parameter.
At large $N$ the absolute value of the order parameter
stays intact across the wall, while $\int dz\, ({\partial\alpha}/{\partial z})
=2\pi /N$. 
Thus, the ``axion''-wall coupling  scales as $N^0$. Note that the sign of the
coupling depends on whether we cross the wall from left to right or
from right to left. This is why the  exchange 
of the dilaton between two parallel BPS walls leads
to the wall attraction while that of the axion leads to repulsion.

\vspace{1mm}

Since the ``dilaton/axion'' coupling to the wall
$\sim N^0$, we
 get no force at the level $N^1$ for free.
The underlying reason is that the BPS domain wall tension scales not as $N^2$,
as would be
 natural for  solitons, but as $N^1$ --- the D brane type of behavior.

\vspace{1mm}

Moreover, we want to say
 that $\sigma$ and $\eta$ give contributions which are exactly
equal in absolute values (at the level $N^0$) but are opposite in signs.
This requires the degeneracy of their masses  ---
which is certainly the case in \None SYM --- and their couplings to the walls
 (except for the relative sign).
Taking the right-hand side of Eq. (\ref{dva})
literally we get the 
``axion''-wall coupling in the form,
\beq
T\,\eta/f + O(1/N)\,,
\label{tri}
\eeq
i.e. the same as in Eq. (\ref{odin}).

\vspace{1mm}

At finite $N$ the wall thickness which scales as \cite{Gabadadze:1999pp}
$(N\Lambda)^{-1}$ is finite too.  We have
$(2+1)$-dimensional supersymmetry on the wall world volume,
which places scalars and pseudoscalars into distinct
(non-degenerate) supermultiplets \cite{AV}. At $N=\infty$
the wall thickness vanishes and it is natural that 
in this limit the wall  coupling involves
the lowest component of $(3+1)$-dimensional
chiral superfield which has the form $\sigma + i\eta$.
This component is coupled to the $(1/2,\,1/2)$
central charge ${\cal Z}$
as ${\cal Z}\exp (\sigma + i\eta) +$ h.c.\footnote{If the BPS wall in \None
gluodynamics becomes an ordinary D-brane at $N=\infty$,
then it must support a massless U(1) gauge field. The gauge field in 1+2 dimensions
is equivalent to a (pseudo)scalar field with the $S_1$ target space \cite{polyakov}.
The above argument suggests an answer to a question
formulated in Ref.~\cite{AV}. Namely, the U(1)
gauge field on the world volume of the domain wall was shown to be described
by the Lagrangian
$$
{\cal L}_{1+2} = -\frac{1}{4e^2}\, F_{mn} F^{mn}
+\frac{N}{16\pi}\, F_{mn}\, A_k\, \varepsilon^{mnk}+\mbox{ferm. terms}.
$$
The Chern-Simons term makes the $A$ field massive, $m_A = N\,e^2/(4\pi)$,
non-degenerate with the translational modulus. The scaling law of $m_A$
depends on that of $e^2$. We suggest that at $N\to\infty$
the degeneracy is restored, i.e. $e^2\sim \Lambda N^{-2}$, so
that $m_A\sim \Lambda N^{-1}\to 0$ at $N\to\infty$.
}

\vspace{1mm}

Thus, the  wall ``R-R charge''
is due to the axion-like nature of $\eta$.  
The ``NS-NS charge'' is due to the wall tension.
 Both charges are indeed equal 
and scale as $N^0\, \Lambda^2$. This guarantees that at order
$N^0$ there is no force. The $\sigma$ and $\eta$ coupling to the wall
split at order $N^{-1}$.

\vspace{1mm}

Needless to say, the very same arguments
can be repeated {\em verbatim}
in the orientifold theories. In the limit $N\to\infty$
the degeneracy of the $\sigma$ and $\eta$ masses holds, and so does
the degeneracy of  the wall ``NS-NS and R-R charges.'' The reason
why 
the couplings to the wall (associated
with the trace of the energy-momentum tensor for the ``dilaton'' and
 the axial charge for the ``axion'') are the same is that these 
charges and couplings are inherited
from the parent supersymmetric theory at  $N\to\infty$, see \cite{Armoni:2003gp}.
For what follows it will be useful to note that
the $\sigma$ exchange alone
(before cancellation) generates the interaction potential between the walls
(per unit area)
\beq
\frac{V}{A} \sim N^0\,\Lambda^3 \, e^{-mZ}\,.
\label{intpotu}
\eeq
This scaling law is in agreement with the string theory expectations.

\vspace{1mm}

Now let us see how this picture is implemented in string theory.
 From the open string standpoint 
 the result of the zero force  is
 natural. This is the Casimir force between the walls. Since the 
 UV degrees of freedom are bose-fermi degenerate the vacuum energy
 and, hence, the Casimir force is zero.

\begin{figure}[H]
  \begin{center}
\mbox{\kern-0.5cm
\epsfig{file=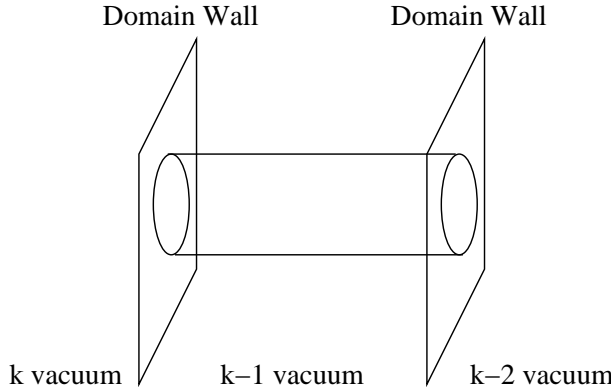,width=8.0true cm,angle=0}}
  \end{center}
\caption{The annulus diagram for the orientifold field theory. The
D branes are domain walls. Closed strings are bosonic glueballs and
open strings are UV degrees of freedom.}
\label{dw}
\end{figure}

In the gauge theory the closed
 strings are the glueballs of the field theory \cite{Witten:1998zw,Csaki:1998qr}. 
Let us consider
 the large separation case, first. At the lowest level we 
 have a massive scalar and a pseudoscalar (we assume a mass gap).
 These two exactly degenerate states correspond to the ``dilaton''
 and ``axion'' (RR 0-form of type IIB/0B). They are expected to become
 massive when the theory is defined on curved background
 \cite{Witten:1998zw}. The type IIB/0B action
 contains the couplings
\beq
 e^{-\Phi}\, {\rm tr}\, F^2 + C\, {\rm tr}\, F \tilde F,
\label{one}
 \eeq
 where $\Phi$ denotes the dilaton and $C$ the R-R zero-form. In addition
 we have a coupling of the graviton and the R-R four-form
\beq
 \eta ^{\mu \rho} h ^{\nu \lambda }\, {\rm tr}\, F_{\mu \nu} F_{\rho
 \lambda} +
 C ^{\mu \nu \rho \lambda }\, {\rm tr}\, F_{\mu \nu} F_{\rho
 \lambda}\,.
\label{two}
 \eeq
In the gauge theory  the ``graviton'' (tensor meson)
 and the four-form are heavier
 glueballs since they carry higher spins than the dilaton and the zero-form.
Similarly, the whole tower
 of degenerate bosonic hadrons of the ``orientifold field theory''
 should correspond to NS-NS and R-R fields of type II/ 0 string theory. 
 This gives
 us a new picture of why the force between domain walls is zero
 in terms of the glueball exchanges: even-parity glueballs lead to
 an attractive force between the walls whereas odd-parity glueballs
 lead to repulsion. The sum of the two is exactly zero
at the leading $N^0$ order.
 The $1/N$ force between the walls is related to
 a possible non-vanishing force between parallel D branes in curved space
 at the order $O( g^2_{\rm st})$.

\vspace{1mm}

 We can also exploit the above picture to estimate the potential between
 a wall and a anti-wall. This configuration is not BPS and a non-zero 
force is expected at the leading $N^0$ order. 
Ar large separations, the force is controlled by
 an exchange of the lowest massive closed strings. These are the
 dilaton and the 0-form. Now, their contributions add
up.  We get an attractive potential
as indicated in Eq. (\ref{intpotu}).

\section{Vanishing of the Vacuum Energy in \boldmath{\None} 
SYM and Orientifold Theory}
\label{VVESOT}

\noindent

One of the surprising results of our previous work~\cite{Armoni:2003gp}
is that the 
$N^2$ part of the vacuum energy density vanishes
 in the ``orientifold field
theory.'' While this result makes sense from the UV point of view,
 where we have bose-fermi degeneracy, it looks 
rather ``mysterious'' from
 the IR point of view, since 
at the level of the composite color-singlet states
we have only bosonic degrees of freedom.
Indeed, since at large $N$ we have free bosons, it is legitimate
to sum the bosonic contributions 
to the vacuum energy density as follows:
\beq
\sum _n \sum _{\vec k} {1\over 2} \sqrt {{\vec k}^2 + M_n^2}, 
\label{ve}
\eeq
where $M_n$ are the hadron masses. The paradox arise since the sum runs over
positive contributions. How can positive contributions sum to zero?

\vspace{1mm}

Before we present our solution, we would like to make two comments.
First, the sum \eqref{ve} is not well defined since the expected
 Regge trajectory is not bounded from above and therefore a
regularization is needed. Second, in the above sum \eqref{ve} 
the $N$ dependence of each individual mode is $N^0$. The expected
UV $N^2$ dependence of the vacuum energy is hidden in the sum over
hadronic modes. Thus, though formally \eqref{ve} represents the
vacuum energy density of the theory, it is not the most efficient
way to calculate it. Below, we present an alternative way of calculation
of the vacuum energy density
--- which explains naturally the vanishing result. 

\vspace{1mm}

Let us consider the contribution to the cosmological constant
from the open string sector. At large $N$ it is dominated
 by the annulus diagram where each boundary consists of $N$ D branes
and a summation over the various D branes is assumed. 
The M\"obius and Klein-bottle as well as higher-genus amplitudes
 are suppressed at large $N$. It is not surprising that the
cosmological constant vanishes, as we have $N^2$ bosons (NS open
strings) and $N^2$ fermions (Ramond fermionic open strings).

\vspace{1mm}

The annulus diagram has another interpretation. It represents
the force between the D branes. The force is mediated by
bosonic closed strings. In a SUSY setup (the type II string),
D branes are BPS objects --- hence,  the  zero force. 
As has been discussed above, the balance, at large separation, is achieved
in this case due to cancellation between the dilaton, the graviton
and the massless R-R forms.

\vspace{1mm}

It is interesting that the force between parallel selfdual D 
branes is zero
also in type-0 string theory~\cite{Klebanov:1998yy,Angelantonj:1999qg}
\beq
{\cal A}= N^2 (V_8-S_8) \equiv 0\,.
\eeq
 This is due to the underlying SUSY on
 the world sheet. The mechanism is exactly as in the type II case:
 the R-R modes cancel the contributions of the NS-NS modes.  Note
that since we are interested only in the planar gauge theory, 
we can restrict ourselves to $g_{st}=0$ on the string theory side.
Therefore, higher-genus amplitudes are irrelevant for our discussion.
At this level the relevant type-0 amplitudes, as well as the bosonic
spectrum, are identical to the type
II ones, in a not too surprising similarity with the situation in the
large $N$ dual gauge theories (the type-0 string becomes, in a sense, 
supersymmeric at the tree level). 
In particular, the induced dilaton tadpole and cosmological constant
are irrelevant, and, thus,  the background inherited from
the supersymmetric theory remains intact.
 
\vspace{1mm}

The vanishing of the annulus diagram
 leads to an explanation of the ``mysterious''
 vanishing of the cosmological constant in the orientifold field
 theory:  if one
 views the hadrons (in the spirit of the AdS/CFT) as closed strings, the
degenerate bosonic spectrum is the reason behind the vanishing result
for both wall-wall interaction and the cosmological constant.

\vspace{1mm}

We hasten to add that though the mechanism
is similar, there is a difference between the two cases: the wall-wall
interaction involves the force between ``D2'' branes (wrapped D5 branes), whereas the
 vanishing cosmological constant involves the force between ``D3''
branes. The two sorts of branes are not necessarily the same --- it
 depends on the specific realization. However, from the bulk point of
 view, the mechanism is identical. The only requirement is the degeneracy
 of the NS-NS and the R-R tower and their couplings to the branes. 

\vspace{1mm}

The difference between string theory and field theory is that in
string theory the force between D branes, from the closed string
standpoint, is related to the contribution to the cosmological
constant from the open string sector. However, closed strings and
open strings are independent degrees of freedom,
 and so one has to add the contribution of the
 closed strings to the cosmological constant.
 In the gauge theory string
picture, the closed strings are simply hadrons made
out of the constituent open strings --- the gluons and quarks.
Therefore, the value of the cosmological constant can be determined by
either iltraviolet (UV)  or infrared (IR) degrees of freedom. It is the
same quantity.

\vspace{1mm}

Below we will present a purely field-theoretic consideration
which will, hopefully, make transparent the issue of  the vinishing of the vacuum
energy density ${\cal E}$ (at level $N^2$) in the orientifold theory. 
Usually it is believed that one needs full supersymmetry
to guarantee that ${\cal E}=0$. It turns out that a milder requirement
--- the degeneracy between scalar and pseudoscalar glueballs/mesons ---
does the same job. Of course, in supersymmetric theories this degeneracy
is automatic. The orientifold field theory is the first example
where it takes place (to leading order in $1/N$) without full supersymmetry.

To begin with, we will outline some general relations relevant to 
${\cal E}$ which are valid in any gauge theory with no mass scale
other than the dynamically generated $\Lambda$.
The vacuum energy density ${\cal E}$ is defined through
the trace of the energy-momentum tensor,
\begin{eqnarray}
{\cal E} & =& \frac{1}{4}\,
\left\langle \theta ^\mu _\mu
\right\rangle  = \frac{1}{4}\, \int DA\, D\Psi\,\,  \theta ^\mu _\mu\,\, \exp (iS)\,,
\nonumber\\[3mm]
 \theta ^\mu _\mu & =& -\frac{3N}{32\pi^2} \, F^2\,,
\label{traceT}
\end{eqnarray}
where
\beq
F^2 \equiv F_{\mu\nu}^a\,  F^{\mu\nu ,a}\,, \qquad F\tilde F\equiv 
 F_{\mu\nu}^a\,  \tilde{F}^{\mu\nu ,a}\,.
\eeq
The second line in Eq. (\ref{traceT}) is exact in \None gluodynamics and is valid
up to $1/N$ corrections in the orientifold theory.

\vspace{1mm}

Now, we
 use an old trick \cite{Novikov:xj} to express the trace of the
energy-momentum tensor in terms of  a two-point function. The idea is to vary
both sides of Eq.~\eqref{traceT} with respect to $1/g^2$
invoking the fact that the only dimensional parameter
of the theory, $\Lambda$, exponentially depends on $1/g^2$. In this
way one  obtains~\cite{Novikov:xj} 
\begin{eqnarray}
{\cal E} & =& -i\int\, d^4x\,\left\langle {\rm vac}\left|
T\left\{ \frac{1}{4}\,\theta ^\mu _\mu (x)\,,\,\, \frac{1}{4}\,\theta ^\nu _\nu (0)
\right\}\right|{\rm vac}\right\rangle_{{\rm conn}}
\nonumber\\[4mm]
 & \equiv & -i\int\, d^4x\,\left\langle \frac{3N}{128\pi^2}\, F^2 (x)\,,
\,\, \frac{3N}{128\pi^2}\, F^2 (0)
\right\rangle_{{\rm conn}}\,.
\label{twopoint}
\end{eqnarray}
The above expression \eqref{twopoint} is formal --- it 
is not well-defined in the ultraviolet.
This is the same divergence that plagues any
calculation of ${\cal E}$ (remember, Eq. (\ref{twopoint})
is general, it is not related to supersymmetry).
In order to give sense to this relation one needs
a particular regularization. For instance, one can always
think of a non-SUSY theory as a SUSY theory with mass terms for
unwanted superpartners (soft SUSY breaking). 

In supersymmetric theories SUSY prompts us
a natural regularization. Indeed, let us consider
the two-point function of the lowest components
of two chiral superfields $D^2 W^2$,
\beq
\langle{\rm tr} D^2 W^2 (x)\,,\,\,{\rm tr} D^2 W^2(0)\rangle\,.
\label{doppyat}
\eeq
Supersymmetry Ward identity tells us that this two-point function
vanishes identically. There are two remarkable facts encoded in Eq.~(\ref{doppyat}).
First, since $D^2 W^2 \propto \left (F^2 + i F\tilde F\right)$,
the vanishing of (\ref{doppyat}) is not due to the boson-fermion
cancellation but, rather, due to the cancellation between
even/odd parity mesons (glueballs).

Second, Eq. (\ref{doppyat}) generalizes Eq.~(\ref{twopoint}),
so that the expression for the vacuum energy density takes the form
\begin{eqnarray}
{\cal E} =  -i\left(  \frac{3N}{128\pi^2}\right)^2 \int\, d^4x\,
\left( \left\langle
 F^2 (x)\,,
\,\, \, F^2 (0)
\right\rangle - \left\langle
F\tilde F (x)\,,
\,\, \, F\tilde F (0)
\right\rangle \right),
\label{ereg}
\end{eqnarray}
where the connected correlators are understood on the
right-hand side. To see that Eq. (\ref{ereg}) is a heir of Eq.
(\ref{twopoint}), please, observe that
\beq
0=  -i\int\, d^4x\,\left\langle \frac{3N}{128\pi^2}\, F\tilde F (x)\,,
\,\, \frac{3N}{128\pi^2}\, F\tilde F (0)
\right\rangle\,.
\label{formzero}
\eeq
The vanishing in Eq. (\ref{formzero}) is due to the fact
that $F\tilde F$ is proportional to the divergence of the axial current $a_\mu$
both in
\None gluodynamics and in the orientifold theory.

It is absolutely  clear that this regularization
works perfectly in the orientifold theories (at level $N^2$).
Indeed, 
the part of the two-point function  that involves $F^2$ is saturated at $N\to
\infty$ by
the propagators of the glueballs with the even parity. Similarly, the part
 that involves $F\tilde F$
is saturated by the odd parity glueballs. We then get
\beq
{\cal E} = \sum _{\rm even\,\, parity} {\lambda _n ^2 \over M_n ^2}
-\sum _{\rm odd\,\, parity} {\lambda _n ^2 \over M_n ^2}\,,\qquad \lambda_n^2\sim N^2
\quad \mbox{for all $n$}\,, 
\label{T-glueballs}
\eeq
where $\lambda _n$ are the couplings to $T_\mu^\mu$ and $\partial^\mu a_\mu$, respectively, and
$M_n$ are the glueballs masses.  Clearly, if  the masses and the
couplings of the glueballs are even/odd-parity degenerate, as 
is the case in \None gluodynamics {\em and} in the
large-$N$ orientifold field theory,
 ${\cal E}$ vanishes.

\vspace{1mm}

In summary, in the ultraviolet calculation the fermi-bose degeneracy
was responsible for the vanishing of the cosmological constant both in 
supersymmetric gluodinamics and in orientifold theory (where 
the cancellation was at level $N^2$). In dealing with ${\cal E}$
a certain regularization procedure is needed. In SUSY it is implicit.
In passing from the UV language to the IR one, we make it explicit
through Eq. (\ref{ereg}). The range of the applicability of the latter is wider
than just SUSY. It is perfectly applicable in the orientifold theories too.

\vspace{1mm}

The
expression \eqref{T-glueballs} is in a remarkable agreement with our
string theory picture. It shows that only bosonic glueballs are
involved and also that the even and odd parity glueballs contribute with the
 opposite signs.

\vspace{1mm}

Perhaps the most interesting lesson from this picture is that the
cosmological constant can vanish  even though only bosonic IR
degrees of freedom are present in the given gauge theory (at least, to the leading order in
$N$). 
In addition, the ``correct'' calculation of the
cosmological constant in \None SYM, using the IR degrees of freedom,
involves a cancellation among the degenerate bosons and omission
of the fermions!

\section{Non-supersymmetric Parent-Daughter
\\  Pairs}
\label{NSPDP}

\noindent

A simple proliferation of the fermion fields in the form of ``flavors''
leads to non-supersymmetric parent-daughter pairs.
This was first mentioned in Ref. \cite{Gorsky:2002wt}
in the context of \ztwo orbifolds. The planar equivalence here holds perturbatively
\cite{Bershadsky:1998mb,Bershadsky:1998cb} but most likely fails
at the nonperturbative level.
If we use, as a starting point, \None gluodynamics and its orientifold,
the planar equivalence has solid chances to hold
nonperturbatively.

Non-supersymmetric planar-equivalent pairs were mentioned in passing in 
Ref.~\cite{Armoni:2003gp}. For instance, gauge theories
with one and the same number of Dirac fermions either in the antisymmetric two-index
or symmetric two-index representations are planar equivalent.
Now we would like to discuss in more
detail parent-daughter 
pairs which are obtained from \None gluodynamics and its orientifold
by introducing fermion replicas. Thus, as previously,
the parent and daughter theories share one and the same gauge group,
U($N$), and one and the same gauge coupling.
The two respective fermion sectors are:

\vspace{1mm}

\noindent
(i) $k$ species of the Weyl fermions in the adjoint,
to be  denoted  as $\left(\lambda^A\right)^i_j$;

and 

\noindent
(ii)  $k$ species of the Dirac fermions
$\left( \Psi^A\right)_{[ij]}$. 

\vspace{1mm}

Here $i,\,j$ are (anti)fundamental indices
running $i,\,j=1,2,...,N$ while $A$ is the flavor index running
$A=1,2,..,k$. Note that $k\leq 5$. Otherwise we loose asymptotic freedom.
Each Dirac fermion is equivalent to two Weyl fermions,
$$
\Psi_{[ij]} \to \{ \eta_{[ij]}\,,\,\,\, \xi^{[ij]}\}\,.
$$

Of course, since both theories are non-supersymmetric,  predictive
power is significantly reduced compared to the case  of
a SUSY parent. 
Still, one can benefit from the comparison of both theories, in particular,
 the Goldstone meson
sectors.
Let us start with the case (ii), $k$ species of 
$ \Psi_{[ij]}$. Since the fermion fields are Dirac and in the complex 
representation of the gauge group, the theory has the same (non-anomalous) chiral symmetry
as QCD with $k$ flavors, namely,
SU($k)_L\times\mbox{SU}(k)_R$.  Various arguments tell us \cite{mac,triangle,rm}
that the pattern of the chiral symmetry breaking is the same as in QCD too,
namely
\beq
\mbox{SU}(k)_L\times\mbox{SU}(k)_R \to \mbox{SU}(k)_V\,.
\label{pcsbone}
\eeq
The only distinction is that in QCD the constant $f$ scales as $\sqrt{N}\,\Lambda$
while in our case its scaling law is ${N}\,\Lambda$. Moreover, the coefficient $n$
in front of the Wess-Zumino-Novikov-Witten term $\Gamma$ (see e.g. Ref.~\cite{Witten:tw})
equals $N$ in QCD and $(1/2) N(N-1)$ in the case at hand (see below).

All axial (non-anomalous) currents are spontaneously broken, giving rise
to $k^2 -1$ Goldstone mesons, ``pions.'' 
Some of them ---  those coupled to the  axial currents that can be elevated
from the daughter theory (ii)  to the parent theory (i) --- 
persist  in the parent theory (i), where the  fermion fields belong to the real
representation,
with the same coupling to the corresponding axial currents. This is
because of the planar equivalence of two theories.
(Remember, currents  with the structure $\bar\xi \xi -\bar\eta \eta$
cannot be elevated from (ii) to (i).)

It is 
not difficult to count the number of the axial currents that are elevated from (ii) to (i):
 there are
$(1/2)k(k-1)$ off-diagonal currents of the type $\bar\xi^A \xi^B +\bar\eta^A \eta^B$
($A\neq B$) plus all $k-1$ diagonal axial currents of the type
$\bar\xi^A \xi^A +\bar\eta^A \eta^A$ (no summation over $A$).
Altogether we get 
$$
\frac{k(k+1)}{2}-1
$$
Goldstone mesons. This obviously corresponds to the following pattern
of the chiral symmetry breaking:
\beq
\mbox{SU}(k) \to \mbox{SO}(k)\,,
\label{pcstwo}
\eeq
with the Goldstone mesons in the symmetric two-index representation of
$ \mbox{SO}(k)$. The pattern of the chiral symmetry breaking for the quarks
belonging to a real representation of the gauge group
indicated in Eq.~(\ref{pcstwo}) was advocated many times
in the literature \cite{mac,triangle,rm,Witten:tx}, but no 
complete proof was ever given.

We conclude this section by a brief comment on the topological properties
of the corresponding chiral Lagrangian
and how they match the underlying gauge field theory expectations.
The theory (i) is expected to be confining
and support flux tubes --- fundamental color charges cannot be screened.
On the other hand, we do not expect stable baryons with mass growing with N.
Composite color-singlet states of gluon and $\left(\lambda^A\right)^i_j$ form
baryons with $M\sim N^0$.

At the same time, the theory (ii) does {\em not} have baryons with
$M\sim N^0$. Here the baryon masses grow with $N$.
The theory is expected to be confining too,
but two flux tubes 
(each attached to a color source in the fundamental representation)
can be screened by $\left( \Psi^A\right)_{[ij]}$.

As was suggested in Refs.~\cite{Witten:tw,Witten:tx},
at large $N$ one can try to identify baryons with the 
Skyrmions supported by the corresponding chiral Lagrangians.
Since
\begin{eqnarray}
&&\pi_3\left\{\mbox{SU}(k)/\mbox{O}(k)
\right\} = Z_4\,\,\,\mbox{at}\,\,\, k=3, \quad \pi_3\left\{\mbox{SU}(k)/\mbox{O}(k)
\right\} = Z_2\,\,\,\mbox{at}\,\,\, k\geq 4;\nonumber\\[2mm]
&&\pi_3\left\{\mbox{SU}(k)
\right\} = Z\,\,\,\mbox{at all}\,\,\, k 
\end{eqnarray}
both theories, (i) and (ii),  yield Skyrmions
with $M_{\rm Skyrme}\sim N^2$, albeit the theory (ii) has a richer spectrum.
The above scaling law,
$M_{\rm Skyrme}\sim N^2$,  is due to the fact that
$f$ scales as $ N$ in the theories under consideration.

 Skyrmion statistics is determined by
the (quantized) factor in front of the Wess-Zumino-Novikov-Witten term,
\beq
(-1)^{N(N-1)/2}\,.
\label{stat}
\eeq
It has half-integer spin provided that $N(N-1)/2$ is odd,
i.e. $N= 4p+2$ or $N= 4p+3$ where $p$ is an integer.
In both cases one can construct, in the microscopic theory,
interpolating baryon currents with an odd number of constituents
scaling as $N$. Why then the Skyrmion  mass scales  as $N^2$?
A possible explanation is as follows.
For quarks in the the fundamental representation of
SU($N$) the color wave function is antisymmetric,
which allows them all to be in the $S$ wave in the coordinate space.
With antisymmetric two-index spinor fields
the color wave function is symmetric, which would require them to 
occupy orbits with angular momentum up to $\sim N$.
Then the scaling law $M_{\rm Skyrme}\sim N^2$ seems natural.

Since $\pi_2\left\{\mbox{SU}(k)
\right\} = 0$ the chiral sector of the theory (ii) does not support flux tubes.
Albeit  disappointing, such a situation was anticipated by Witten
\cite{Witten:tx} who noted that topology of the full space 
of the large $N$ theory need not coincide with topology of 
its Goldstone sector.

In light of this remark we can understand the complete failure
of the Skyrmion description of the theory (i). In particular,
since $\pi_2\left\{\mbox{SU}(k)/\mbox{O}(k)
\right\}$ $ = Z_2$ at $ k\geq 3$
we get flux tubes in the chiral theory, while we do not expect them
in the microscopic theory. Moreover, stable Skyrmions of 
the chiral sector should become unstable in the full
theory.

\section{Conclusions}

\noindent

In this work we tried to further develop a parallel
between \None gluody\-namics and its 
non-supersymmetric orientifold daughter on the one hand,
and string/D brane paradigm, on the other.
We discussed forces between two 
BPS domain walls in field theory terms
and established contact with the string theory description.
In the latter, there is a well-known cancellation between
NS-NS and R-R interactions. In field theory terms
this cancellation manifest itself as follows:
the exchange of a composite dilaton coupled
to the domain wall is canceled (at leading order)
by that of a composite axion.
The string-theory interpretation allows us to establish
a one-to-one relation between vanishing force and 
vanishing cosmological constant.
Thus, the question ``who is responsible for the
vanishing cosmological constant in non-supersymmetric orientifold
field theory?'' gets a rather unexpected answer ---
the degeneracy of the even-odd parity composite mesons.

In the last part of the paper we 
discuss  ``flavor proliferation''
as a device allowing one to get
 planar equivalent pairs
in which both theories in the parent-daughter pair are
non-supersymmetric, starting from the original pair ---
\None gluodynamics and its orientifold daughter.
Although we loose predictive power based on supersymmetry
of the parent, some  predictions survive.
In particular, we compare Goldstone meson sectors,
and obtain consequences for the
patterns of the chiral symmetry breaking.

\Acknowledgements

We are grateful  to  I. Brunner, G. Gabadadze,
 I. Kogan,  W. Lerche, A. Loewy, R. Rabadan,
J. Sonnenschein,
and G. Veneziano for fruitful discussions.
Special thanks go to  A. Gorsky and A. Ritz 
for insightful communications.
The work of M.S. is supported in part  by DOE grant DE-FG02-94ER408.

\end{document}